\documentstyle[aps,prl,epsf,twocolumn]{revtex}
\begin{document}
\twocolumn[\hsize\textwidth\columnwidth\hsize\csname
@twocolumnfalse\endcsname
\title{Detecting the Kondo screening cloud around a quantum dot}
\author{Ian Affleck$^{1,2}$ and Pascal Simon$^{2}$}
\address{$^1$Canadian Institute for Advanced
Research  and $^2$Department of Physics and Astronomy, University
of British Columbia, Vancouver, B.C., Canada, V6T 1Z1}
 \maketitle
\begin{abstract}
A fundamental prediction of scaling theories of the Kondo effect
is the screening of an impurity spin by a  cloud of electrons
spread out over a mesoscopic distance.  This cloud has
never been observed experimentally.  Recently, aspects of  the
Kondo effect have been observed in experiments on quantum dots
embedded in quantum wires.  Since the length of the wire may be of
order the size of the screening cloud, such systems provide an
ideal opportunity to observe it.  We point out that persistent
current measurements in a closed ring provide a conceptually
simple way of detecting this fundamental length scale.
\end{abstract}
]
The almost trivial looking problem of a single quantum spin
interacting with a gas of otherwise non-interacting electrons has
attracted much attention from condensed matter theorists for
decades \cite{Hewson}. In particular, it played an important role in the
development of scaling and renormalization group (RG) ideas.
While the ``bare'' dimensionless coupling constant, $\lambda$ is
small, the low energy behavior of the system is described by an
infinite effective Kondo coupling. This implies that a magnetic
impurity makes a large contribution to the resistivity of the
metal at low temperatures, corresponding to the ``unitary limit''.
The characteristic energy scale at which the effective coupling
constant becomes large defines the Kondo temperature:
\begin{equation}
T_K\approx De^{-1/\lambda},\label{TK}\end{equation} where $D$ is
the bandwidth and we have set Boltzmann's constant to one. 
Associated with this energy scale is a fundamental
length scale:
\begin{equation}
\xi_K= \hbar v_F/T_K,\end{equation} where $v_F$ is the Fermi velocity.
For a typical metal exhibiting the Kondo effect this length scale
can be of order .1 microns or larger.  A simple picture of the
meaning of this length scale is that a single electron occupying a
wave-function extended over this distance forms a spin singlet
with the impurity spin (which is assumed to have $S=1/2$ here).
Physical quantities such as the Knight shift can be expressed in
terms of universal scaling functions of $r/\xi_K$ where $r$ is the
distance from the impurity \cite{Sorensen,Barzykin}.  Recent numerical 
simulations \cite{Sorensen} have
explicitly confirmed such scaling behavior. Nevertheless, no
experiment has ever directly detected this fundamental length
scale.  The reason for this is perhaps mainly that this scale is
too large. 

 While the original Kondo effect led to a large {\it resistivity} at low
 $T$, a quite different manifestation of the same physics was
 predicted in a slightly different experimental configuration,
 corresponding to a large {\it transmission} coefficient \cite{Glazman,Ng}.  A
 quantum dot can act like an S=1/2 impurity due to the Coulomb
 blockade.  By adjusting a gate voltage on the dot, it can be
 arranged that the groundstate of the dot has an odd number of
 electrons, with an unpaired spin pointing up or down, and that
 the energy cost to add or remove an electron is relatively large.
  When this quantum dot is connected to leads with weak tunneling
  matrix elements, the transmission coefficient of the dot is
  small at most temperatures.  This system is essentially
  equivalent to the Kondo model with the dimensionless coupling
  constant being of order $t'^2 /\epsilon_0D$ where $t'$ is the tunneling
  matrix element, $\epsilon_0$ is
  the energy to add or remove an electron from the dot and $D$ is the
  bandwidth of the leads.
  Now the growth of this effective coupling constant at low
  temperatures is manifested by the dot exhibiting a transmission
  coefficient close to 1 at temperatures below $T_K$.  Recent
  developments in nanotechnology have made it possible to observe
  various aspects of this ``transmission Kondo effect'' in the last
  few years \cite{Goldhaber,Cronenwett,Simmel,Wiel}.

  Since Kondo temperatures in these experiments are generally
  considerably smaller than $1^0K$ (and can be tuned via the gate
  voltage, $\epsilon_0$) the Kondo length scale is expected to be
  of order 1 micron or larger.  Again, the simple picture is that
  an electron in the leads, occupying a wave-function spread out
  over this length scale, screens the spin of the quantum dot.
  This new experimental realization of the Kondo effect seems to
  offer new opportunities to measure the screening cloud.  In
    some of  these experiments \cite{Wiel}, the quantum dot is connected to
  thin leads (quantum wires) of length of order 1 micron, which,
  in turn are connected to macroscopic leads further from the dot.
   One might hope to see a change in behavior when the length of
   the quantum wire leads becomes comparable to $\xi_K$. (The dot
  itself is generally much {\it smaller} than the screening cloud.)
Indeed, this notion of suppression of the Kondo effect by a finite 
size is central to much of the theoretical work on the problem; see 
especially the nice interpretation of Wilson's numerical 
renormalization group calculations by Nozi\`eres \cite{Nozieres}.
     However,
   the situation in these experiments  
is complicated by the fact that the screening
   cloud may also occupy the macroscopic parts of the leads.
   While devices can be envisioned\cite{Fran} that might get around
   this problem, our goal here is to study the conceptually
   simplest way of probing screening cloud physics in such
   systems.

   We consider a quantum dot in a closed mesoscopic circular
   ring \cite{Buttiker,Ferrari,Kang}. 
   In this geometry the screening cloud is ``trapped'' in
   the ring and cannot escape into macroscopic leads.  The
   experimental challenge is to measure the transmission
   amplitude through the dot without  attaching leads to the ring.
   This could be done by measuring the
   persistent current in the ring induced by a magnetic flux.
   Such persistent current experiments have been performed
   recently on micron sized rings not containing a quantum
dot \cite{Chand,Mailly}.
    We shall assume here that the ring
   contains no other impurities besides the quantum dot.  It is
   encouraging in this regard to note that recent experiments,
   with macroscopic leads attached, observed perfect conductance
   ($2e^2/h$) through a quantum dot \cite{Wiel}.  We only consider 
$T=0$ here (effectively $T<<T_K$, $T<<\hbar v_F/L$) although our 
results could be extended straightforwardly to higher $T$. 
 One might expect that the
   persistent current, as a function of the flux, $\Phi$ penetrating the
   ring, will be much different when the screening cloud is much smaller
   than the circumference, $L$, of the ring than when it is much
   larger.  We will argue below that, when $\xi_K<<L$, the
   persistent current is that of a perfect ring with no impurity,
   of order of magnitude $j\propto ev_F/L$.
   On the other hand, when $\xi_K >>L$, $jL$ becomes much
   smaller, vanishing as a power of the bare Kondo coupling.  In
   general, $j(\Phi )L$ is a universal scaling function of
   $\xi_K/L$, in the usual scaling limit of the Kondo model (i.e.
   at small Kondo coupling and large ring size compared to the lattice
 constant).  The functional
   form of $j(\Phi )$ crosses over from a saw-tooth for $\xi_K<<L$
   to a sine function at $\xi_K>>L$.

   We begin with the standard tight-binding Anderson model for the
   quantum dot-quantum wire system \cite{Glazman,Ng}, $H=H_0+H_{int}$ with
\begin{eqnarray}
   H_0&=&-t\sum_{j\leq -2}(c^\dagger_jc_{j+1}+h.c.)-t\sum_{j\geq 1}
   (c^\dagger_jc_{j+1}+h.c.)
   \nonumber \\
H_{int} &=& -t'[c^\dagger_0(c_{-1}+c_1)+h.c.]
   +\epsilon_0c^\dagger_0c_0
   +Un_{0\uparrow}n_{0\downarrow}.\label{Hand}\end{eqnarray}
   We then pass to the Kondo
   limit, $t'<<-\epsilon_0$, $U+\epsilon_0$, 
 where the dot is singly occupied and virtual tunneling to the
   neighboring sites (at $j=\pm 1$) lead to a spin-exchange
   interaction:
   \begin{equation}
   H_{int}=J(c^\dagger_{-1}+c^\dagger_1)
   {\vec \sigma \over 2}(c_{-1}+c_1)\cdot \vec S.\label{Hkondo}\end{equation}
The Kondo coupling is
$J=2t'^2[-\epsilon_0^{-1}+(U+\epsilon_0)^{-1}]$. The dimensionless
Kondo coupling appearing in Eq. (\ref{TK}) is $\lambda =
4J\sin^2k_F/\pi v_F$, where $k_F$ is the Fermi momentum.  The zero
temperature persistent current is given by $j=-(e/\hbar )
dE_0/d\alpha$ where $E_0$ is the groundstate energy, and a
magnetic flux, $\Phi = (\hbar c/e)\alpha$ is applied to the ring,
corresponding to modifying phases of hopping terms so that the
 sum of phases is $\alpha$.  We consider a ring of $L$ sites.
To calculate $j$ in the large $L$ limit, we may 
linearize the dispersion relation around the Fermi surface, leading 
to simple formulas for the electron propogator in the time domain and 
thus facilitating perturbation theory in $\lambda$.
 Ignoring corrections down by $1/L$, and working to $O(\lambda^2)$ for an
 even number of electrons, $N$ and to $O(\lambda^3)$ for odd $N$,
 we find:
 \begin{eqnarray}
j_e(\alpha ) &=& {3\pi v_F e\over 4  L}\{ [\sin \tilde \alpha
[\lambda + \lambda^2\ln (Lc)]\nonumber \\
&& +(1/4 +\ln 2)\lambda^2\sin
2\tilde \alpha \} + O(\lambda^3)\ \  \nonumber \\
j_o(\alpha )&=&{3\pi v_F e\over 16  L}\sin 2\alpha
[\lambda^2+2\lambda^3\ln (Lc')]+O(\lambda^4),\label{jpert}\end{eqnarray}
for $N$ even and odd respectively,  where $c$ and $c'$ are
constants of O(1) which we have not determined and:
\begin{eqnarray}
\tilde \alpha &=& \alpha \ \  (N/2\  \hbox{even})\nonumber \\
\tilde \alpha &=& \alpha + \pi \  \  (N/2\
\hbox{odd}).\label{tildedef}\end{eqnarray}

Importantly, $j(\alpha ,\lambda ,L)L$,
 to the order we have worked, is a function
only of $\alpha$ and the renormalized Kondo coupling at
scale $L$: $\lambda_{eff}(L)=\lambda + \lambda^2 \ln L + \ldots$.
That this should be true exactly, in the scaling limit, 
 follows from standard RG
arguments.  (See, for example \cite{Barzykin}.)
  Since the current is conserved it may be calculated at
an arbitrary point in the ring, far from the quantum dot.
Therefore it has vanishing anomalous dimension since the
interactions all take place near the origin and anomalous
dimensions vanish for all operators far from the impurity.
Therefore the dimensionless quantity $Lj$ can only depend on the
effective coupling at scale $L$ or equivalently on the ratio
$\xi_K/L$.  In particular, this implies that the perturbative
result becomes valid at small $L/\xi_K$ where $\lambda_{eff}(L)$
is small.  The corrections to this scaling form are suppressed by 
factors of $a/L$, where $a$ is the lattice constant (which we 
have set to 1).

To calculate $j$ at large $L/\xi_K$, we may use the fact that
$\lambda_{eff}\to \infty$ and that $j$ is a (universal)
characteristic of the infrared fixed point.  Thus we may obtain it
from the Kondo Hamiltonian of Eq. (\ref{Hkondo}) by setting $J\to
\infty$.  In this limit one electron is trapped in the symmetric
orbital on  sites $\pm 1$, $(1>+|-1>)$.  The low energy effective
Hamiltonian is simply a non-interacting tight-binding model
\begin{eqnarray}
H_{low}&=&-t\sum_{j\leq -3}(c^\dagger_jc_{j+1}+h.c.)
-t\sum_{j\geq
2}(c^\dagger_jc_{j+1}+h.c.)\nonumber \\
&& -{t\over \sqrt{2}}(-c^\dagger_{-2}c_a
+c^\dagger_ac_2+h.c.),\label{Hlow}\end{eqnarray} where 
$c_a\equiv (c_1-c_{-1})/\sqrt{2}$.
  This model exhibits resonant transmission at $k=\pi /2$,
corresponding to half-filling, due to particle-hole symmetry.  The
persistent current in such a non-interacting, scattering model, is
completely determined by the transmission amplitude at the Fermi
surface \cite{Gogolin}.  Thus, in the half-filled case, 
we obtain the persistent
current for an ideal ring.  We should take into account an
effective shift of $\alpha$ by $\pi$ due to the reversed sign for
the hopping term between sites $-2$ and $0$ in Eq. (\ref{Hlow})
and also the fact that the number of low energy electrons is
$N-2$. These two effects on the persistent current cancel so that 
it is exactly the same as in original model of Eq. (\ref{Hand}) 
with $U=0$, $t'=t$.  Thus
we obtain, after simply adding the contributions of spin up and down 
electrons, the persistent current for $L>>\xi_K$:
\begin{eqnarray}
j_e(\alpha ) &=&-{2ev_F\over \pi L}[\tilde \alpha  -\pi ],\ \  (N
\ \hbox{even})\nonumber \\
j_o(\alpha ) &=& -{ev_F\over \pi L}([\alpha ] + [\alpha - \pi ])\
\ (N \ \hbox{odd}).\label{jstrong}
\end{eqnarray}
It can easily be proven, for $L>>1$ and arbitrary $\xi_K/L$, 
that $j_e$ for the two cases
of $N/2$ even or odd are related by a $\pi $ shift of $\alpha$,
i.e. that a single functon $j_e(\tilde \alpha )$ describes both
cases.  Thus the persistent current, for $L>>1$ and $\lambda <<1$
is given by two universal scaling functions $j_eL(L/\xi_K,\tilde
\alpha )$ and $j_oL(L/\xi_K,\alpha )$.  We have calculated these
functions in the limits of small and large arguments in Eqs.
(\ref{jpert}) and (\ref{jstrong}).  They are shown in FIG. 1.
\begin{figure}[ht]
\epsfxsize=2.5 in\centerline{\epsffile{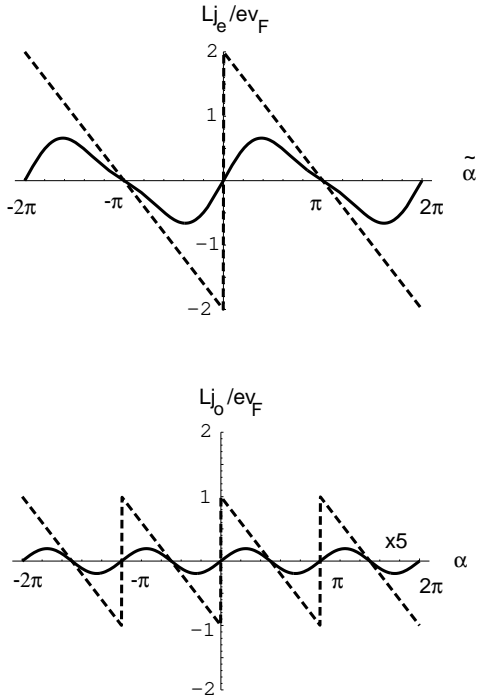}}
\caption{Persistent current vs. flux for an even or odd number of 
electrons for $\xi_K/L\approx 50$ (solid line) and for $\xi_K/L<<1$
 (dashed line).  $j_o$ 
is multiplied $\times 5$ at $\xi_K/L=50$ for visibility.  The 
solid lines are obtained from Eq. (\ref{jpert}) using the effective 
coupling $\lambda (L) \approx 1/\ln (\xi_K/L)$.}
\end{figure} 
We expect the current at intermediate $\xi_K/L$ to interpolate 
smoothly between these limits.  
Note that the persistent current has the same sign, for all $\alpha$,
at both small and large $\xi_K/L$, as in the non-interacting limit 
of the Anderson model, thus respecting the Leggett
conjecture. Also note that $j_o(\alpha )$ has period $\pi$, while
 $j_e(\tilde \alpha )$ has period $2\pi$ for both large and small $\xi_K/L$. 
(We suspect that these properties
 are true in general, but have not been able to 
prove this.)  There is a large parity effect,
especially at large $\xi_K/L$, where the current is much smaller for $N$ odd.

In fact, the $J\to \infty$ limit of Eq. (\ref{Hlow}), 
only gives perfect transmission
for $k_F=\pi /2$, at half-filling; in general the transmission 
amplitude is $T(k)=\sin^2k$ for $H_{low}$.  However, it is known that
particle-hole symmetry breaking is an exactly marginal
perturbation in the Kondo problem so that for small bare Kondo
coupling, the effective particle-hole symmetry breaking remains
small in the low energy limit. (See, for example \onlinecite{Barzykin}
.)  Thus we expect the persistent
current to depend only weakly on electron density and gate voltage, 
for all $\xi_K/L$, in the Kondo regime for small bare Kondo coupling. 
This situation changes considerably if we include screened Coulomb
interactions in the ring.  Then particle-hole symmetry breaking
becomes relevant (for repulsive interactions) and the persistent
current only achieves its ideal value for $L>>\xi_K$ at a special
resonant value of the gate voltage, $\epsilon_0$ \cite{Kane,Gogolin}.

If we take non-symmetric tunneling amplitudes from the wire to the
dot, $t_l'\neq t_r'$, then the persistent current at $L>>\xi_K$ is
reduced by a factor proportional to 
 $[2t_r't_l'/(t_r'^2+t_l'^2)]^2$ but still exhibits
scaling behavior.

Real quantum wires have several active channels (i.e. transverse
sub-bands).  However, it is reasonable to expect that one channel
will have a stronger tunneling amplitude to the dot than the
others.  The RG equations, to third order, for the multi-channel
Kondo problem are:
\begin{equation}
d\lambda_i/d\ln l = \lambda_i^2-(1/2)\lambda_i\sum_j\lambda_j^2.
\label{RG3}\end{equation}
We see that if all but one of the couplings, $\lambda_1$, are small, while 
$\lambda_1$ is larger and positive, then we may approximate the equation 
for the small couplings by only the second term in Eq. (\ref{RG3}), keeping 
only the $\lambda_i\lambda_1^2$ term.  This equation then predicts 
that all the small couplings shrink.  Meanwhile, the larger
coupling grows. It can be easily seen from this equation that this 
behavior is generic.
The impurity spin is screened by an electron from 
the most strongly coupled channel and the other channels decouple at 
low energies. Thus,
for $L>>\xi_K$ we expect the single channel result of Eq.
(\ref{jstrong}) to still apply.

We have also studied the case of a ``side-coupled'' quantum dot.
In this case, the electrons can only hop from the dot to site $0$
on the chain.  The corresponding Kondo Hamiltonian is:
\begin{equation}
H_R =-t\sum_{j=-\infty}^\infty (c^\dagger_jc_{j+1}+h.c.)
+Jc^\dagger_{0}
   {\vec \sigma \over 2}c_{0}\cdot \vec S.\end{equation}
   This Hamiltonian is well-known to exhibit perfect {\it
   reflection} ($\pi /2$ phase shift in the even channel) at low
   energies.  Thus the transmission amplitude vanishes so that the
   persistent current is zero for $L>>\xi_K$.  On the other hand,
   for small $J$, the persistent current is nearly that of an
   ideal ring.  This is the inverse of the situation for the
   embedded dot, discussed above.  We note however, that this
   simple model give a much less realistic description of a 
real experiment in the side-coupled
   case.  In this case all of the channels should have significant
   transmission past the dot, for small $J$.  With increasing $L$
   only one channel gets a reduced current as a result of
   screening the impurity spin.  Furthermore, if the quantum dot
   is represented by more than one side coupled site, at large
   $J$, the screening cloud can form entirely off the ring, with
   no reduction of the persistent current.  In this situation it
   is far from clear what the behavior will be for small bare
   coupling and large $L/\xi_K$.

   The persistent current for the embedded dot was studied
   in \onlinecite{Kang}, by solving some approximate
   self-consistent equations.  The results were much different than ours.
   In particular, the current was predicted to be much smaller for
   $N$ odd than for $N$ even {\it for all} $L/\xi_K$. We find this result only 
in the perturbative regime, $\xi_K>>L$.  This appears to indicate
   that these self-consistent equations are not sensitive to the
   infrared divergences at large $L$ related to the RG flow.  (We have 
similar disagreements on the side-coupled case also studied using 
the same method \cite{Cho}.)

   The side-coupled dot was also studied in \onlinecite{Eckle}.
   The opposite conclusion to ours was
   reached that the ring would exhibit a perfect persistent
   current, rather than zero current, at large $L/\xi_K$.  We
   believe that this incorrect result arose from a
   misinterpretation of some excitation parameters in the Bethe
   ansatz solution and an invalid general formula for the persistent current 
    which treats incorrectly the
   contribution to the current from electrons below the Fermi
   surface. 

In conclusion, we have shown that the persistent current through a 
quantum dot in a mesoscopic closed ring depends strongly on the 
ratio of the screening cloud size to the ring circumference.  This 
provides hope that the elusive Kondo screening cloud may eventually 
be directly measured.

\acknowledgements I. A. thanks Silvano De Franceschi and the organizers 
of the NATO Workshop on Size-Dependent Magnetic Scattering 
 (P\'ecs, Hungary, May 2000) for stimulating his interest in
this subject.  This research was supported in part by NSERC of Canada.

\end{document}